\documentstyle[aps,eqsecnum,preprint,floats,graphicx]{revtex}

\newcommand{\bra}[1]{\langle #1 |}
\newcommand{\ket}[1]{| #1 \rangle}
\newcommand{\Ham}{{\cal H}}

\tighten
\begin{document}

\draft
\preprint{\parbox{4.5cm}{
UMIST/Phys/TP/98-6%\\nucl-th/98xxxxx
}} 
\title{Collective coordinates, shape transitions and shape coexistence:
a microscopic approach}
\author{Takashi~Nakatsukasa\footnote{
Electronic address: T.Nakatsukasa@umist.ac.uk}
and
Niels~R.~Walet\footnote{
Electronic address: Niels.Walet@umist.ac.uk}
}
\address{Department of Physics, UMIST, P.O.Box 88,
Manchester M60 1QD, UK}

\maketitle

\begin{abstract}
We investigate a description of shape-mixing and shape-transitions
using collective coordinates. To that end we apply a theory of
adiabatic large-amplitude motion to a simplified nuclear shell-model,
where the approximate results can be contrasted with exact
diagonalisations. We find excellent agreement for different regimes,
and contrast the results with those from a more standard calculation
using a quadrupole constraint. We show that the method employed in
this work selects diabatic (crossing) potential energy curves where
these are appropriate, and discuss the implications for a microscopic
study of shape coexistence.
\end{abstract}


%\narrowtext

%%%%%%%%%%%%%%%%%%%%%%%%%%%%%%%%%%%%%%%%%%%%%%%%%%%%%%%%%%%%%%%%%%%%%%%%
\section{Introduction}

In order to describe processes in nuclei involving large excursions
from equilibrium, such as shape-coexistence or fission, we cannot use
the small-amplitude harmonic expansion about a stationary mean-field
state provided by the random-phase approximation (RPA). Various
methods exist to deal with large-amplitude motion, many of which are
reviewed in Ref.~\cite{KWD91}. This reference also sets out the basic
formalism on which our work is based.

We have recently embarked on a study into the properties of collective
motion in systems with pairing. In a first application we have
analysed properties of collective motion in a semi-microscopic model
of nucleons interacting through a pairing force, coupled to a single harmonic
variable giving a macroscopic description of the remaining degrees
of freedom\cite{NW98}. We have used the local-harmonic approximation,
which is equivalent to the generalised valley approximation for a
single collective coordinate \cite{KWD91}, to analyse the collective
motion. To our surprise it has turned out that the system
automatically selects either diabatic or adiabatic collective surfaces
according to the strength of the pairing interaction.  However, since
this model is not fully microscopic, we feel that it would be
beneficial to study a fully microscopic Hamiltonian.  This does not
mean that we wish to fully ignore the success of the unified model by
Bohr and Mottelson, which indicates that the semi-macroscopic approach
describes many nuclear phenomena quite well, but rather that we wish
to understand such behaviour from a microscopic viewpoint.  We thus
feel that it is desirable to test the methodology for a fully microscopic
Hamiltonian which is able to describe nuclear systems from vibrational
to deformed.

To this end we investigate the collective motion in a microscopic model
which describes a system of nucleons interacting through a simplified
version of the pairing-plus-quadruple force \cite{PSS80}.
Although the Hamiltonian has a very simple form,
we shall see that the model can reproduce the qualitative features of
many kinds of interesting situations observed in real nuclei.
In this work we shall concentrate on the spherical-to-deformed
transition, and nuclei with shape coexistence, where more than
one equilibrium shape plays a role.

In the case of a single-$j$ shell the model Hamiltonian is built from
the generators of an $o(4)$ algebra, which makes exact diagonalisation
feasible. The model has been originally developed to describe
$K^\pi=0^+$ excitations in deformed nuclei \cite{PSS80}, and has also
been used as a test-bed for various methods used in the calculation of
collective excitations such as the boson expansion method
\cite{Mat82}, the selfconsistent collective coordinate method
\cite{Mat86}, and a semiclassical method \cite{SM88}.  The model can
be generalised to multiple shells, where it has been used to investigate
shape-coexistence phenomena \cite{FMM91}. Finally, a similar model has
been used to study the collective mass parameter in finite
superconducting systems \cite{AB88}.

Although the low-lying spectra in nuclei are mostly dominated by the
quadrupole phonon ($J^\pi=2^+$) excitations, the anharmonicity is very
important for many nuclei, especially in a shape-transition region,
where the nature of the ground-state changes rapidly with particle
number.  For instance, the even-even Sm isotopes show a typical
example of the spherical to deformed shape transition in which the
spectrum shows a strong anharmonicity between the spherical ($N\leq
84$) and deformed ($N\geq 90$) nuclei, especially for $^{148,150}$Sm.
These phenomena are primarily related to the competition between the
monopole and quadrupole interactions among the valence particles
outside a closed core.  The pairing-plus-quadrupole model, originally
proposed by Bohr and Mottelson, was designed to describe this
competition and is quite well able to reproduce the most important
aspects of the experimental data (see \cite{PPQQ} and references
therein).  Later the boson expansion method has been applied to the
same model (with an additional quadrupole pairing interaction) for the
description of the shape transition in the Sm isotopes \cite{KT72},
which shows excellent agreement with the experimental data.  Since the
$O(4)$ model is very similar to the pairing-plus-quadrupole model, it
would be of significant interest to see whether our method of large
amplitude collective motion is able to properly describe the shape
transition phenomena in this exactly solvable model.

The importance of shape-coexistence in nuclear physics can be seen
from the multitude of theoretical approaches and the amount of
experimental data as compiled in the review paper Ref.~\cite{Woo92}.
An important example can be found in even semimagic Sn and Pb
isotopes, where the ground states are spherical.  However, deformed
excited $J^\pi = 0^+$ states have been observed at low-excitation
energies in many of these nuclei.  These excited states are regarded
as states associated with proton two-particle-two-hole (2p-2h)
excitations across the closed shell.  Using the Nilsson picture, which
shows down-sloping single-particle levels above the proton closed
shell, and up-sloping levels below it, it is possible to assign a
configuration of two particles lying on down-sloping levels and two
holes on up-sloping levels.  The configuration-constrained
Nilsson-Strutinsky calculations as performed by Bengtsson and
Nazarewicz \cite{BN89} have suggested that the diabatic
potential-energy curve obtained by switching off the interaction
between the 2p-2h and the ground-state (0p-0h) configuration gives
more accurate picture than the conventional adiabatic potential
energy.  This question, whether the nuclear collective potential is
adiabatic or diabatic, is quite old, and was originally raised by Hill
and Wheeler \cite{HW53}.  We have shown in our previous work
\cite{NW98} that, using a method to self-consistently determine
collective coordinates, the system itself selects either an adiabatic
or a diabatic collective path according to the properties of the
interaction.  It is our aim to investigate in the $O(4)$ model
whether the method is able to provide us with useful information about
shape mixing, and to test whether it makes useful predictions whether
the collective potential energy is diabatic or adiabatic.

In Sec.~\ref{sec:formalism}, the theory of adiabatic large amplitude
collective motion is briefly reviewed.
A problem peculiar to the model under consideration, related to the
special character of the zero modes, is dealt with by a prescription to
remove the zero-mode degrees of freedom.  In Sec.~\ref{sec:O4_model},
the $O(4)$ model in a single-$j$ shell is described, and we discuss
the applications of our methodology to the model.  In
Sec.~\ref{sec:multi_O4}, a generalisation of the $O(4)$ model to
multiple shells is introduced.  We then investigate the large
amplitude collective motion for a set of parameters which describe
transitional (from vibrational to deformed) nuclei.  In
Sec.~\ref{sec:two_comp_O4}, we use a proton-neutron form of the
Hamiltonian, which can be used to describe shape-coexistence. We show
how our method selects a diabatic or adiabatic potential energy curve.
Finally, we give some conclusion and present an outlook in
Sec.~\ref{sec:conclusion}.

%%%%%%%%%%%%%%%% Review of LACM %%%%%%%%%%%%%%%%%%%%%%%%%%%%%%%%%%
\section{formalism}
\label{sec:formalism}

\subsection{Review of methodology for the local harmonic approximation}

We briefly review the local harmonic approximation (LHA), as a method
in adiabatic large amplitude collective motion (ALACM). A full
discussion of the method can be found in Ref.~\cite{KWD91}.  As usual
in our work, we use the convention that the repeated appearance of the
same symbol ($\alpha,\beta,\cdots; i,j,\cdots$) as an upper and lower
index denotes a sum over the relevant index over all allowed values.
We also use the convention that a comma in a lower index denotes the
derivative with respect to a coordinate, thus $F_{,\alpha} = \partial
F/\partial \xi^\alpha$.

Our approach to
large amplitude collective motion
is applicable for  classical Hamiltonian systems
which have a kinetic energy quadratic in momentum only.
Since most systems, especially the mean-field problems of nuclear physics,
do not satisfy this requirement, we are forced to truncate the full
Hamiltonian $\Ham(\xi,\pi)$ to second order,
\begin{equation}
\label{H_ad}
\Ham_{ad}(\xi,\pi) = \frac{1}{2} %\sum_{\alpha\beta}
         B^{\alpha \beta} \pi_\alpha \pi_\beta + V(\xi) \ ,
\hspace{1cm} \alpha, \beta = 1,\cdots,n \ .
\end{equation}
Here the mass tensor $B^{\alpha \beta}$ depends in general on
the coordinates $\xi^\alpha$ and is defined by
expansion of the Hamiltonian to second order in momenta,
\begin{equation}
B^{\alpha \beta} = \left.
        \frac{\partial^2 \Ham}{\partial\pi_\alpha \partial\pi_\beta}
        \right|_{\pi=0} \ .
\end{equation}
Thus all terms of more than quadratic order in momenta are neglected.
This is only possible when the  higher-order terms are small,
which is definitely true in the small (zero) velocity limit. It is
in this sense that the theory may be regarded as an {\it adiabatic} theory.
The tensor $B_{\alpha \beta}$, which is defined as the inverse of
$B^{\alpha \beta}$
($B^{\alpha \gamma} B_{\gamma \beta} = \delta^\alpha_\beta$),
plays the role of metric tensor in the Riemannian formulation of
the local harmonic approximation (LHA) below.

Collective coordinates $x^i$ and intrinsic (non-collective)
coordinates $x^a$ which are decoupled from each other,
are assumed to be reached  by making a point transformation,
conserving the quadratic nature of Eq. (\ref{H_ad}),
\begin{eqnarray}
x^i &=& f^i(\xi)  \quad (i=1,\cdots, K) \ ,\\
x^a &=& f^a(\xi)  \quad (a=K+1,\cdots, n) \ .
\end{eqnarray}
In this section, we use symbols ($\alpha,\beta,\cdots$) as indices of the
original coordinates, ($\mu,\nu,\cdots$) for the new coordinates after the
transformation, ($i,j,\cdots$) for collective coordinates and
($a,b,\cdots$) for intrinsic coordinates.  The requirement that the
motion is exactly restricted to a collective subspace $\Sigma$ (defined
by $x^a=0$), yields three conditions,
\begin{eqnarray}
\label{mass_diagonal}
\bar{B}^{ai} = 0 \ ,
\label{real_force_cond}
\bar{V}_{,a} = 0 \ ,\\
\label{geometrical_force_cond}
\bar{B}^{ij}_{,a} = 0 \ .
\end{eqnarray}
These three conditions are only satisfied for exactly decoupled
collective motion, a rare occurrence indeed.  
It is therefore practical to combine 
these three decoupling conditions into two sets of conditions, that
are satisfied even when no exact collective subspace exists. The conditions
chosen are those that determine the bottom of a valley in the potential 
landscape, which is, under certain conditions, an approximation to
a decoupled subspace. The quality of this decoupling can be measured,
see below. In this work we prefer to work with the LHA.
In the default case of a single collective coordinate
only ($K=1$), the basic equations of this formalism can be written as
\begin{eqnarray}
\label{LHA_1}
V_{,\alpha} = \lambda f^1_{,\alpha} \ ,\\
\label{LHA_2}
B^{\beta\gamma} V_{;\alpha\gamma} f^1_{,\beta} = \omega^2 f^1_{,\alpha}\ .
\end{eqnarray}
The second of these equation is the local harmonic (or local RPA)
equation from which the method derives its name.
The covariant derivative (denoted by $;$) in the left-hand side
of Eq. (\ref{LHA_2}) is defined by
\begin{equation}
V_{;\alpha\beta} \equiv V_{,\alpha\beta}
               - \Gamma^\gamma_{\alpha\beta} V_{,\gamma} \ ,
\end{equation}
with the affine connection $\Gamma$ defined in the standard way in
terms of the metric tensor $B_{\alpha\beta}$ as
\begin{equation}
\label{affine_1}
\Gamma^\alpha_{\beta\gamma} = \frac{1}{2}
   B^{\alpha\delta} \left( B_{\delta\beta, \gamma}
                        + B_{\delta\gamma , \beta}
                        - B_{\beta\gamma , \delta} \right) \ .
\end{equation}

The equations (\ref{LHA_1}) and (\ref{LHA_2}) must be solved as a
self-consistent pair, except at stationary points, where the RPA
(\ref{LHA_2}) is independent of the force equation (\ref{LHA_1}).
This allows us to bootstrap our way up from such a stationary point.  This
procedure constructs a path by finding successive points at which an
eigenvector $f^1_{,\alpha}$ of the covariant RPA equation
(\ref{LHA_2}) satisfies the force condition (\ref{LHA_1}) at the same
time.  It is worth noting that since the equations at every point can
be solved independently, no computational error is accumulated while
calculating the collective path, which is a problem with, e.g., the
formalism of Goeke and Reinhard \cite{GR}.

The quality of decoupling can be measured by comparing two different
collective mass parameters that can be calculated in the theory.  If
we calculate the derivatives $d\xi^\alpha/dx^1$ in terms of the
tangents of the path, we find
\begin{equation}
\check{B}_{11} =
    \frac{d\xi^\alpha}{dx^1} B_{\alpha\beta} \frac{d\xi^\beta}{dx^1} \ .
\end{equation}
The other mass parameter can be obtained by using the eigenvectors
$f^1_{,\alpha}$ obtained from the covariant RPA equation,
\begin{equation}
\bar{B}^{11} = f^1_{,\alpha} B^{\alpha\beta} f^1_{,\beta} \ .
\end{equation}
This is equal to $(\bar{B}_{11})^{-1}$ if the decoupling is exact.
Therefore, we define the decoupling measure $D$ as
\begin{equation}
\label{decoupling}
D = \left(\check{B}_{11} \right) \bar{B}^{11} - 1 \ .
\end{equation}
The size of this measure $D$ indicates the quality of decoupling. The smaller
its value, the better the decoupling.

%%%%%%%%%%%%%%%%%%%%%%%%%%%%%%%%%%%%%%%%%%%%%%%%%%%%%%%%%%%%%%%%%%%%%%%%
\subsection{Removal of Spurious Modes}
\label{sec:prescription}

In this section we discuss how to treat the spurious modes.
A typical example is given by the Nambu-Goldstone (NG) mode associated with
the violation of particle-number conservation.
We have presented a method adding a constraint to the original LHA
formalism, in order to
find a collective subspace orthogonal to the NG modes \cite{NW98}.
However, in the model to be discussed here  this method does not work because
of a divergence problem associated with the fact that the model
has an exact zero mass parameter, $\det(B^{\alpha\beta})=0$.
Instead, in this paper, we choose to remove the NG degrees of freedom
explicitly.

For the models to be discussed in the following sections,
the modes associated with a change in average particle number
are given by a linear combination of coordinates:
\begin{equation}
\tilde{f}^{\sc ng}(\xi) = \sum_{\alpha=1}^n c_\alpha \xi^\alpha \ ,
\end{equation}
where $c_\alpha$ is a constant. The problem is that this
mode leads to a zero eigenvalue of the mass,  \begin{equation}
B^{\alpha\beta} \tilde{f}^{\sc ng}_{,\beta}
          = B^{\alpha\beta} c_\beta = 0 \ .
\end{equation}
This means that we cannot invert the mass matrix. The only sensible
way to deal with this is to remove these degrees of freedom from our
space, by defining a new set of coordinates,
$\tilde{\xi}^{\mu} = \tilde{f}^\mu(\xi)$.
These are required to satisfy
\begin{eqnarray}
\label{new_co_1}
B^{\alpha\beta} \tilde{f}^\mu_{,\beta} \neq 0 \ ,\quad&\mbox{ for }&
\forall\alpha\mbox{ and }\mu=1,\cdots,n-M \ ,\\
\label{new_co_2}
B^{\alpha\beta} \tilde{f}^\mu_{,\beta} = 0 \ ,\quad&\mbox{ for }&
\forall\alpha\mbox{ and }\mu=n-M+1,\cdots,n \ ,
\end{eqnarray}
where we assume that there are $M$ NG modes ($\mu > n-M$).
Then, we may formulate the LHA
in the space of $n-M$ dimension, $\{\tilde{\xi}^\mu\}_{\mu=1,\cdots,n-M}$,
in which $\det(B^{\mu\nu}) \neq 0$.
\begin{eqnarray}
{\cal M}^\nu_\mu f^i_{,\nu} = (\omega^i)^2 f^i_{,\mu} \ ,\\
{\cal M}^\nu_\mu \equiv \tilde{V}^{,\nu}_{;\mu}
                  = \tilde{B}^{\nu\nu'}\tilde{V}_{;\nu'\mu} \ ,
\end{eqnarray}
where indices $\mu,\nu,\cdots$ are running only from 1 to $n-M$.
Our aim is to provide a feasible method to calculate this LHA,
namely to calculate the mass parameter $\tilde{B}^{\nu\nu'}$,
potential $\tilde{V}(\tilde{\xi})$, and their derivatives.

The second equation (\ref{new_co_2}) determines
tangent vectors of the NG modes.
The rest of coordinates $\tilde{f}^\mu$ for $\mu=1,\cdots,n-M$
are arbitrary as long as
their derivatives are linearly independent from the others.
%For instance, in the case that we have non-zero $M$-dimensional
%Jacobian for\footnote{One can make this always true by simply changing
%the order of coordinates.}
%\begin{equation}
%\det\left(
%\frac{\partial\tilde{\xi}^\mu}
%     {\partial \xi^\alpha} \right)
%= \det\left( \tilde{f}^\mu_{,\alpha} \right)
%= \det\left( c^\mu_\alpha \right) = \mbox{const}
%\neq 0 \ ,
%\end{equation}
%where $(\mu,\alpha) = n-M+1,\cdots,n$.
%Then we may simply take
%\begin{equation}
%\tilde{\xi}^\mu = \delta^\mu_\alpha \xi^\alpha \ ,
% \quad\quad \tilde{f}^\mu_{,\alpha} = \delta^\mu_\alpha\ ,
% \quad\mbox{ for }
% (\mu,\alpha) = 1,\cdots,n-M \ .
%\end{equation}
The full Jacobian matrix $\tilde{f}^\mu_{,\alpha}$
allows us to define the derivatives of inverse transformation,
$\tilde{g}^\alpha_{,\mu}$ as the inverse of $\tilde f$,
\begin{equation}
\tilde{f}^\mu_{,\beta} \tilde{g}^\beta_{,\nu} = \delta^\mu_\nu
\ ,\quad
\tilde{f}^\mu_{,\beta} \tilde{g}^\alpha_{,\mu} = \delta^\alpha_\beta \ .
\end{equation}
Since all $\tilde{f}^\mu_{,\alpha}$ are constant
(independent of coordinates),
all $\tilde{g}^\alpha_{,\mu}$ are also constant and
the derivatives $\tilde{f}^\mu_{,\alpha\beta}$
(or $\tilde{g}^\mu_{,\alpha\beta}$) 
all vanish.
This implies that within the NG subspace the  connection  vanishes,
$\Gamma = 0$,
and the geometric character of the transformation of any tensor 
is fully determined in the subspace that does not contain the NG modes.
One can use this to calculate the new mass parameter and its derivatives as
\begin{eqnarray}
\tilde{B}^{\mu\nu} &=&
      \tilde{f}^\mu_{,\alpha} B^{\alpha\beta} \tilde{f}^\nu_{,\beta} \ ,\\
\tilde{B}^{\mu\nu}_{,\lambda} &=&
      \tilde{f}^\mu_{,\alpha} B^{\alpha\beta}_{,\gamma}
      \tilde{f}^\nu_{,\beta} \tilde{g}^\gamma_{,\lambda} \ ,
\end{eqnarray}
and the derivatives of potential as
\begin{eqnarray}
\tilde{V}_{,\mu} &=& \tilde{g}^\alpha_{,\mu} V_{,\alpha}\ ,\\
\tilde{V}_{,\mu\nu} &=& \tilde{g}^\alpha_{,\mu} \tilde{g}^\beta_{,\nu}
                      V_{,\alpha\beta}\ .
\end{eqnarray}

%For the case of single NG mode,
%the Jacobian matrices of the transformation and
%of the inverse transformation are given by
%\begin{eqnarray}
%\tilde{f}^\mu_{,\beta}
%   \equiv \frac{\partial \tilde{\xi}^\mu}{\partial \xi^\beta}
%    &=& \left\{\begin{array}{ll}
%      \delta^\mu_\beta\ , &\quad \mbox{for }\mu = 1,\cdots,n-1 \ ,\\
%       c_\beta\ ,         &\quad \mbox{for }\mu = n \ ,
%       \end{array} \right. \\
%\tilde{g}^\alpha_{,\mu}
%   \equiv \frac{\partial \xi^\alpha}{\partial \tilde{\xi}^\mu}
%    &=& \left\{\begin{array}{ll}
%      \delta^\alpha_\mu\ , &\quad \mbox{for }\alpha = 1,\cdots,n-1 \ ,\\
%       \frac{1}{c_n}\left\{\delta^\alpha_\mu (1+c_\mu) - c_\mu\right\}\ ,
%                           &\quad \mbox{for }\alpha = n \ ,
%       \end{array} \right.
%\end{eqnarray}
%where $c_n \neq 0$ is assumed.
%Then, one can apply the normal LHA to the space reduced to $(n-1)$
%dimensions.

%%%%%%%%%%%%%%%%%%%%%%%%%%%%%%%%%%%%%%%%%%%%%%%%%%%%%%%%%%%%%%%%%%%%%%%%
\section{The $O(4)$ model}
\label{sec:O4_model}

We shall first study the properties of the single-shell $O(4)$ model.
We define fermionic  operators
$c^\dagger_{jm}$ and $c_{jm}$ that create or annihilate a particle in
the $J_z=m$ sub-state. In terms of these operators we define four pairing 
($P$, $P^\dagger$, $\tilde P$, and $\tilde P^\dagger$) and two 
multipole operators ($N$ and $Q$) that close under commutation, and 
generate the algebra $o(4)$,
\begin{eqnarray}
P^\dagger &=& \sum_{m>0} c_{jm}^\dagger c_{j\bar m}^\dagger \ , \quad
\tilde{P}^\dagger = \sum_{m>0} \sigma_{jm} c_{jm}^\dagger c_{j\bar m}^\dagger \ ,\\
N         &=& \sum_m c_{jm}^\dagger c_{jm}\ , \quad
Q         = \sum_m \sigma_{jm} c_{jm}^\dagger c_{jm}\ , \quad\\
\sigma_{jm}  &=& \left\{
               \begin{array}{ll}
               +1 & \mbox{for } |m| < \Omega/2\\
               -1 & \mbox{for } |m| \geq \Omega/2
               \end{array} \right. .
\end{eqnarray}
Here we need to require that the pair multiplicity $\Omega = j+1/2$
is an {\em even} integer in order for the algebra to close.  The sign
of $\sigma_{jm}$ is chosen so as to mimic the behaviour of the matrix
elements of the axial quadrupole operator $\bra{jm} r^2 Y_{20}
\ket{jm}$, and we shall call $Q$ the quadrupole operator in the
remainder of this work, even though it does not carry the correct
multipolarity.
%The four operators, $P$, $\tilde{P}$, $N$ and $Q$,
%generate the Lie algebra $o(4)$.

As is well-known, the algebra $o(4)$ is isomorphic to $su(2)\oplus
su(2)$.  This can be made explicit in terms of the quasi-spin operators
\begin{eqnarray}
\label{su2_gen_1}
A_+ &= \frac{1}{2} \left( P^\dagger + \tilde{P}^\dagger \right)
      = A_-^\dagger\ ,
 \quad A_0 =& \frac{1}{4} \left( N + Q - \Omega \right) \ ,\\
\label{su2_gen_2}
B_+ &= \frac{1}{2} \left( P^\dagger - \tilde{P}^\dagger \right)
      = B_-^\dagger\ ,
 \quad B_0 =& \frac{1}{4} \left( N - Q - \Omega \right) \ ,
\end{eqnarray}
which generate two independent $su(2)$ algebras,
\begin{eqnarray}
\left[ A_+ , A_- \right] &=& 2 A_0 \ ,  \quad \left[ B_+ , B_- \right] = 2 B_0 \ ,\\
\left[ A_0 , A_\pm \right] &=& \pm A_\pm \ , \quad  \left[ B_0 , B_\pm \right] = \pm B_\pm \ ,\\
\left[ { A}_\mu , { B}_{\mu'} \right] &=& 0 \ .
\end{eqnarray}

The Hamiltonian of the model is chosen as a simple quadratic form in (some of)
the generators of $o(4)$,
\begin{equation}
H = -G P^\dagger P - \frac{1}{2} \kappa Q^2 \ ,
\label{O4_Hamiltonian}
\end{equation}
and mimics the pairing-plus-quadrupole model that has been such a
successful phenomenological model in heavy nuclei \cite{PPQQ}.  Even
though the Hamiltonian looks simple, it does not have a closed-form
solution (it does not have $O(4)$ dynamical symmetry). Nevertheless a
numerically exact solution for the Hamiltonian (\ref{O4_Hamiltonian})
can be obtained by simple diagonalisation.  To this end one rewrites
the Hamiltonian  in terms of the quasi-spin
operators ${\bf A}$ and ${\bf B}$,
\begin{equation}
\label{H_O4}
H = -G \left( A_+ + B_+ \right) \left( A_- + B_- \right)
    - 2 \kappa \left( A_0 - B_0 \right)^2 \ .
\end{equation}
This Hamiltonian  commutes with the total particle number $N=2(A_0
+ B_0) + \Omega$, and there are no further constants of the motion.
The pairing force tends to align the two quasi-spin vectors ${\bf A}$
and ${\bf B}$, so as to obtain the maximal pairing matrix elements,
while the quadrupole force tends to de-align them (to maximise $(A_0 -
B_0)^2$).  In this picture, the non-integrability of the model, as
well as the physics described, is related to the competition between
the pairing and the quadrupole force. This is identical to a
competition between alignment and de-alignment of the quasi-spins.

For a fixed number of particles $N=2n_0$,
we  construct from the vacuum state $\ket{0}$ all states with
a constant number of generators $A_+$ and $B_+$,
\begin{equation}
\ket{n_0,k_a} = 
\left[ \frac{ \left(\frac{\Omega}{2} - k_a \right)!
     \left(\frac{\Omega}{2} - n_0 + k_a \right)! }
   { \left\{\left(\frac{\Omega}{2}\right)!\right\}^2
   k_a ! \left( n_0 - k_a \right)! } \right]^{1/2}
   A_+^{k_a} B_+^{n_0 - k_a} \ket{0} \ ,\\
\end{equation}
where $0 \leq k_a \leq n_0$.
Finding the eigenvectors of the Hamiltonian now involves a trivial
matrix diagonalisation in this basis of dimension ($n_0 + 1$).

%%%%%%%%%%%%%%%%%%%%%%%%%%%%%%%%%%%%%%%%%%%%%%%%%%%%%%%%%%%%%%%%%%%%%%%%
\subsection{The Coherent-State Representation
and the TDH(F)B Equations of Motion}

The mean-field description of the Hamiltonian (\ref{H_O4}) is most
easily based on the use of a product of $su(2)$ coherent states, one for
the $A_\mu$ sub-algebra, and another for the $B_\mu$ one.  Each of these
states is characterised by a complex variable, $z_a$ and $z_b$
\cite{Per86}.  The time-dependent mean-field dynamics in this
parametrisation is the classical Hamiltonian problem we shall apply
our methodology to.  We can also parametrise the coherent state with
four real angles \cite{SM88,Per86}, 
\begin{eqnarray}
\ket{z_a,z_b} &=& \exp\left[z_a A_+ - z_a^* A_-
                          + z_b B_+ - z_b^* B_- \right] \ket{0} \ ,\\
   &=& \left( \cos\frac{\theta}{2} \cos\frac{\chi}{2} \right)^{\Omega/2}
       \exp\left[ \tan\frac{\theta}{2} \exp(-i\phi) A_+
            + \tan\frac{\chi}{2} \exp(-i\psi) B_+ \right] \ket{0}\ ,
\end{eqnarray}
where we have used
\begin{equation}
z_a = \frac{\theta}{2} \exp(-i\phi) \ ,\quad
z_b = \frac{\chi}{2} \exp(-i\psi) \ .
\end{equation}

The time-dependent Hartree-Fock Bogoliubov (TDHFB) equations are in
this case the classical equations of motion obtained from the
stationary condition of the coherent-state action $\delta S = 0$,
where
\begin{eqnarray}
S[z] &=& \int^t dt\, \bra{z_a,z_b} i \partial_t - H \ket{z_a,z_b} \ ,\\
     &=& \int^t dt\,
          \frac{\Omega}{2} \left( \dot{\phi} \sin^2\frac{\theta}{2}
          + \dot{\psi} \sin^2\frac{\chi}{2} \right)
          - \int^t dt\, \Ham (\theta, \chi; \phi,\psi) \ ,
\end{eqnarray}
and
\begin{equation}
\label{clH_0}
\Ham = \bra{z_a,z_b} H \ket{z_a,z_b} \ .
\end{equation}
In order to facilitate our work we  introduce
real canonical variables $\xi^\alpha$ and  $\pi_\alpha$,
\begin{eqnarray}
\xi^1 &=& \frac{\Omega}{2} \sin^2 |z_a|
           = \frac{\Omega}{2} \sin^2 \frac{\theta}{2}\ , \quad
  \xi^2 = \frac{\Omega}{2} \sin^2 |z_b|
           = \frac{\Omega}{2} \sin^2 \frac{\chi}{2}\ ,  \\
\pi_1 &=& \arg(z_a) =  -\phi\ , \quad 
  \pi_2 = \arg(z_b) =  -\psi\ .
\end{eqnarray}
Since these variables are canonical, the equations of motion are of
Hamiltonian form
\begin{equation}
\dot{\pi}_\alpha = -\frac{\partial \Ham }{\partial \xi^\alpha} \ , \quad
\dot{\xi}^\alpha = \frac{\partial \Ham }{\partial \pi_\alpha} \ ,
\end{equation}
where the classical Hamiltonian (\ref{clH_0}) is the coherent state
expectation value of the Hamiltonian rewritten in terms of canonical
variables,
\begin{eqnarray}
\label{clH_1}
\Ham &=& {\cal H}_P + {\cal H}_Q \ ,\\
\label{clH_P_1}
\Ham_P(\xi,\pi) &=& -\frac{G}{16} \left\{
               \left| \sum_\alpha e^{i\pi_\alpha} S_\alpha \right|^2
              + 32 \Omega^{-1}\sum_\alpha (\xi^\alpha)^2 \right\} \ ,\\
\label{clH_Q_1}
\Ham_Q(\xi) &=& -2\kappa \left\{
               \left( \sum_\alpha (-)^{\alpha+1} \xi_\alpha \right)^2
   +  \Omega^{-1} \sum_\alpha \xi^\alpha (\Omega - \xi^\alpha ) \right\} \ ,\\
S_\alpha &=& 2 \sqrt{2\xi^\alpha(\Omega - 2\xi^\alpha)} \ .
\end{eqnarray}
The terms proportional to $\Omega^{-1}$ in Eqs. (\ref{clH_P_1}) and
(\ref{clH_Q_1}) originate from the exchange (Fock) contributions.

Expanding the Hamiltonian with respect to $\pi$ up to second order,
the mass parameter $B^{\alpha\beta}(\xi)$ and the potential $V(\xi)$,
which both depend only on the coordinates, are defined as
\begin{eqnarray}
\Ham_{\rm ad} &=& \frac{1}{2} \sum_{\alpha\beta}
            B^{\alpha\beta} \pi_\alpha \pi_\beta + V(\xi) \ ,\\
B^{\alpha\beta} &=& \frac{G}{8} 
\left[\delta_{\alpha\beta}
       \left( S_\alpha\sum_\gamma S_\gamma\right) -S_\alpha S_\beta \right]\ ,
         \quad \mbox{(no summation with respect to $\alpha$)} \ ,\\
V(\xi) &=& V_P(\xi) + V_Q(\xi) \ ,\\
\label{V_P_1}
V_P(\xi) &=& \Ham_P(\xi, \pi=0) = -\frac{G}{16} \left\{
               \left( \sum_\alpha S_\alpha \right)^2
              + 32 \Omega^{-1}\sum_\alpha (\xi^\alpha)^2 \right\} \ ,\\
\label{V_Q_1}
V_Q(\xi) &=&  -2\kappa \left\{
               \left( \sum_\alpha (-)^{\alpha+1} \xi_\alpha \right)^2
   +  \Omega^{-1} \sum_\alpha \xi^\alpha (\Omega - \xi^\alpha ) \right\} \ .
\end{eqnarray}
In the rest of this work we shall invoke the Hartree-Bogoliubov approximation,
and we neglect the
exchange terms (Fock terms) in Eqs. (\ref{V_P_1}) and (\ref{V_Q_1}).

%%%%%%%%%%%%%%%%%%%%%%%%%%%%%%%%%%%%%%%%%%%%%%%%%%%%%%%%%%%%%%%%%%%%%%%%
\subsection{Requantisation}

In this section, we discuss the problem of defining a requantisation
procedure and the consequences of the adiabatic truncation with
respect to momentum.  The classical limit of the single-$j$
Hamiltonian has two constants of motion: $\Ham = E$ and $\langle N
\rangle = 2 \sum_\alpha \xi^\alpha = N_0$.  Since the phase space is
four dimensional, this implies the complete integrability of the
system, and there is a two dimensional torus on which all classical
orbits lie.  Due to this special feature of this model, one can apply
the Einstein-Brillouin-Keller (EBK) quantisation condition.  This has
been done in Ref.\ \cite{SM88} and good agreement with the exact
results has been obtained for both energy spectra and transition
amplitudes.  However, it is impossible to extend this
quantisation method to non-integrable systems like the ones we will
discuss in the following sections.  We wish to use the same quantisation
procedure for the simplest form of the model and the more
complicated cases discussed later on, and shall turn to our favourite technique first.

After truncation of the Hamiltonian up to second order in momentum, we
can define a collective Hamiltonian by evaluating its value 
for points on the collective space $\Sigma$ which is
parametrised by $x^1$ and $p^1$ (strictly this is the co-tangent
bundle over $\Sigma$), since we have chosen $x^a=p^a=0;
a=1,\cdots,n-1$,
\begin{eqnarray}
\bar{\Ham}_{\rm col} &=& \left.\bar{\Ham}_{\rm ad}\right|_{\Sigma}
         = \frac{1}{2} \bar{B}^{11}(x^1) p_1^2
            + \bar{V}(x^1) \ ,\\
\bar{B}^{11}(x^1) &=& \sum_{\alpha\beta}
         f^1_{,\alpha} f^1_{,\beta}
         B^{\alpha\beta}\left(\xi^\alpha = g^\alpha(x^1, x^a=0) \right) \ ,\\
\bar{V}(x^1) &=&  V\left(\xi^\alpha = g^\alpha(x^1, x^a=0) \right)\ .
\end{eqnarray}
Since the scale of collective coordinate $x^1$ is arbitrary,
we choose to normalise $f^1_{,\alpha}$ so as to make $\bar{B}^{11} = 1$.
Subsequently, the Hamiltonian $\bar{\Ham}_{\rm col}$
is quantised in this flat space as \cite{NW98}
\begin{equation}
\label{H_col_1}
\hat{H}_{\rm col} = - \frac{1}{2} \frac{d^2}{dx^2} + \bar{V}(x) \ ,
\end{equation}
where we employ the simplification of denoting $(x^1,p_1)$ by $(x,p)$.

In order to evaluate the matrix elements of a one-body operator $F$
(either diagonal or transition matrix elements), we first obtain the
collective classical representation of the operator $F$, which in
keeping with the adiabatic approximation is expanded in powers of
momentum,
\begin{equation}
\bar{\cal F}(x,p) =
\left. {\cal F}(\xi,\pi)\right|_{\Sigma}
   = \left. \bra{z} F \ket{z}\right|_\Sigma
   = \sum_{i=0}^\infty \left. {\cal F}^{(i)}(\xi,\pi) \right|_\Sigma
   = \sum_{i=0}^\infty \bar{\cal F}^{(i)}(x,p) \ .
\end{equation}
Here ${\cal F}^{(i)}$ is the term of $i$-th order in $\pi$.
The function $\bar{\cal F}$ is requantised,  by making the replacement
$\bar{\cal F}(x,p) \rightarrow \bar{F}(x,\frac{d}{dx})$,
at which point one will have to confront the problem of operator
ordering between $x$ and $p$. We shall avoid this problem by
keeping,
invoking once again the assumption of slow collective motion,
only the zeroth order term $\bar{\cal F}^{(0)}$.
It is an interesting question what the effect of higher order
terms will be. This is clearly outside the scope of the present work, and
requires further investigation.
Fortunately, in the current model,
we have no ambiguity for the quadrupole operator $Q$
because ${\cal Q}^{(i)} = 0$ for $i\neq 0$. For convenience we denote the
classical limit of the quadrupole operator by $q$.
The transition matrix elements can thus be calculated by the
one-dimensional integral,
\begin{equation}
\bra{n'} F \ket{n} = \int dx \Psi_{n'}^*(x) \bar{F}^{(0)}(x)
                               \Psi_n(x) \ ,
\end{equation}
where $\Psi_n$ are the eigenfunctions of the collective Hamiltonian
(\ref{H_col_1}),

\begin{figure}[htb]
\centerline{\includegraphics[width=0.9\textwidth]{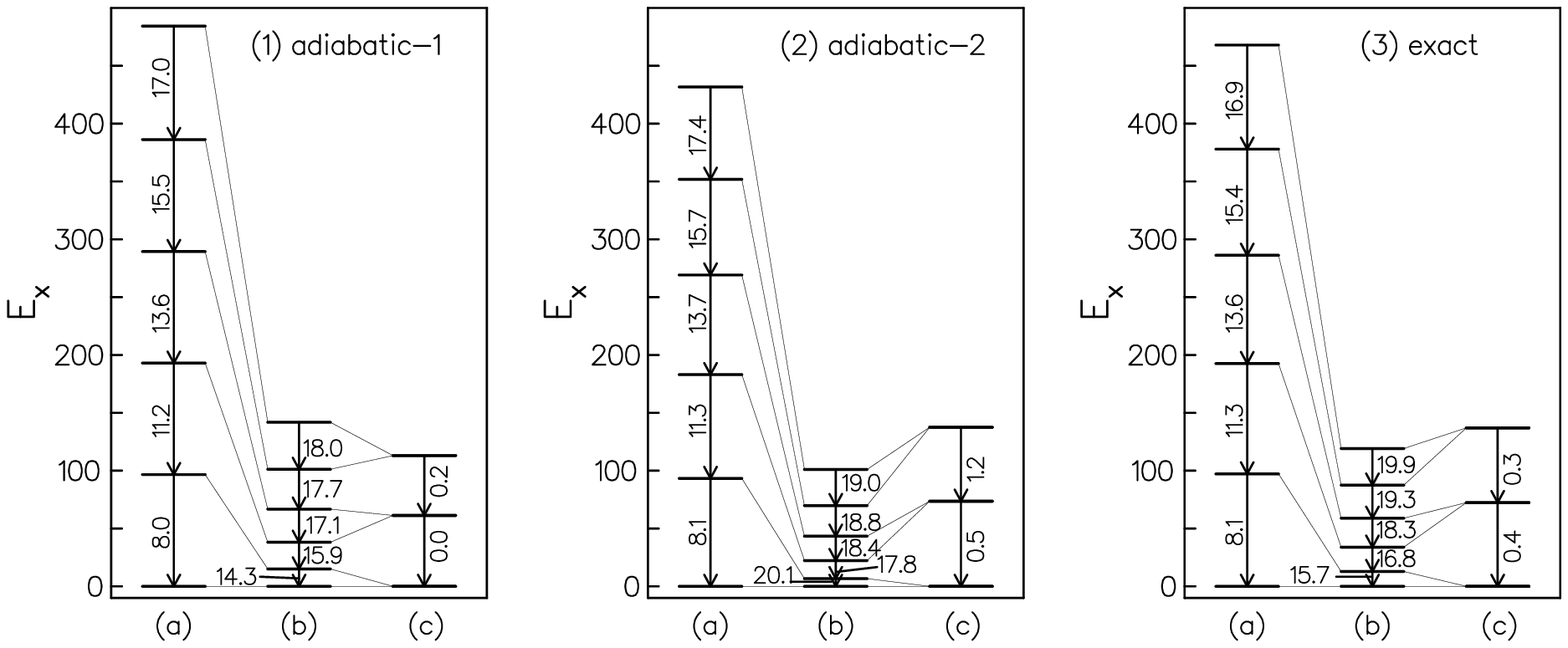}}
\caption{The
excitation energies and transition matrix elements $|\bra{n-1} Q
\ket{n}|$ in the single-$j$ shell $O(4)$ model as a function of the
ratio between the strength of the pairing and that of the
quadrupole force.  The case (a) is a weak quadrupole force, $2\kappa/G=0.079$,
(b) a medium sized one, $2\kappa/G=1.63$ and (c) a very strong one, 
$2\kappa/G=12.7$.
The three panels give our standard adiabatic quantisation (1), the 
results from the ``adiabatic in coordinates'' method (2), 
and the exact results (3).
}\label{fig:singlej-1}
\end{figure}

\begin{figure}[htb]
\centerline{\includegraphics[width=0.5\textwidth]{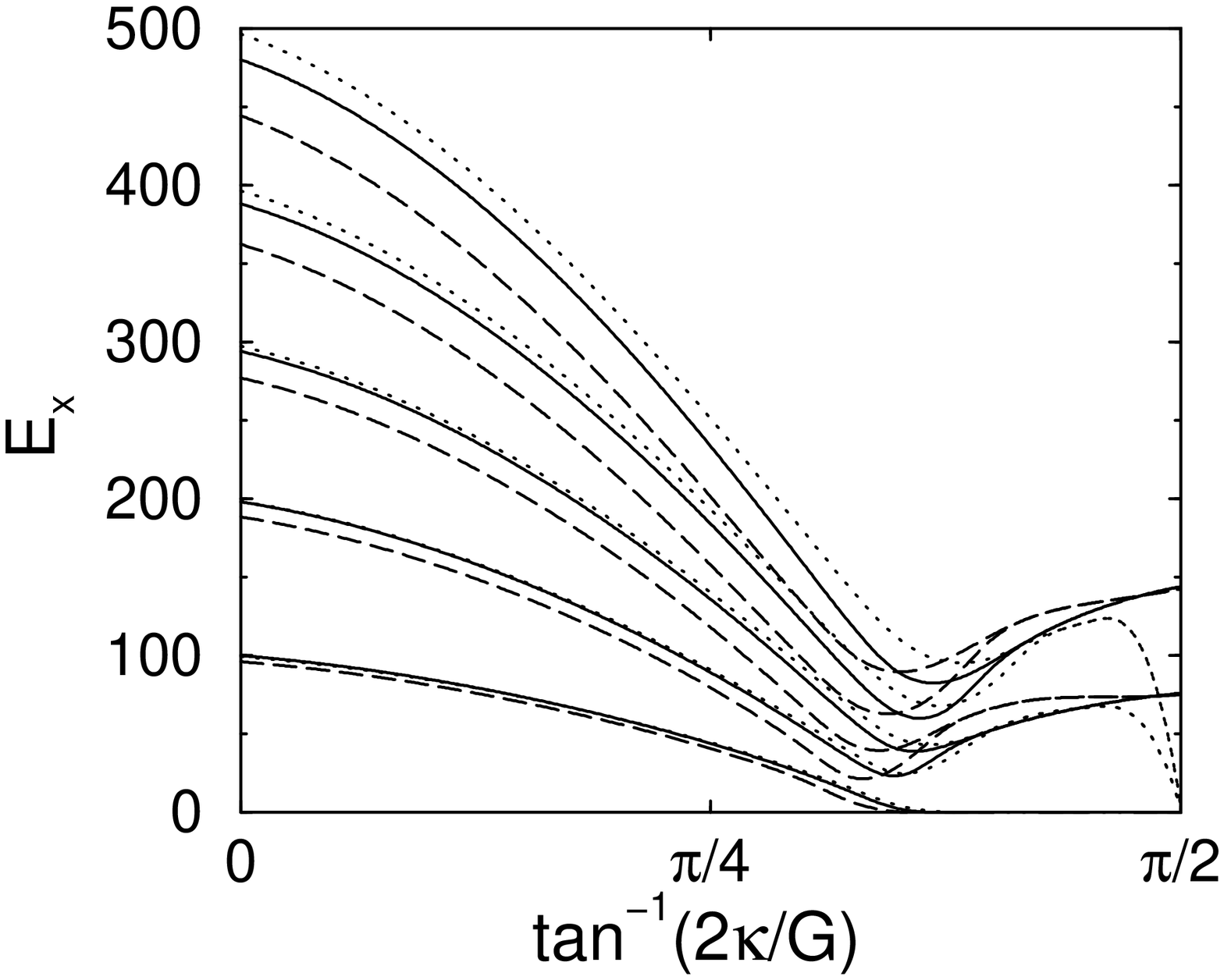}}
\caption{The excitation energies in the single-$j$ shell $O(4)$ model
for 40 particles in a shell with $\Omega=100$ as a function of the
parameter essentially the ratio between the strength of the pairing
and that of the quadrupole force.  The left end corresponds to the
case of pure pairing force and the right end to the pure quadrupole
force.  The  solid lines are the exact results, and the dotted
lines represent the standard requantisation of the adiabatic
collective Hamiltonian. The dashed lines represent the expansion in
terms of coordinates discussed in the text.  }\label{fig:singlej-2}
\end{figure}

From the number of coordinates and momenta found (2+2) we see that the
configuration space of the single-$j$ shell model is two-dimensional.
Since there is a zero mode ($\det(B^{\alpha\beta}) = 0$) corresponding
to the NG mode associated with the particle number violation, one may
obtain a one-dimensional path $\Sigma$ by simply applying the
prescription in Sec.~\ref{sec:prescription} (without the application
of LHA)! Rather than plotting this path we have chosen to represent
the results of requantisation for energies and transition strengths.
These results are presented in Fig. \ref{fig:singlej-1}.  We
obtain good agreement with the exact results over a wide range of
parameters except very close to a pure quadrupole force.  Due to the peculiar
nature of the quadrupole operator the mass parameter goes to zero in this
limit ($G=0$), and there is no
kinetic term.  Thus an eigenstate of Hamiltonian is a coordinate
eigenstate $\ket{x}$ at the same time.  Then, the periodic nature of
the momentum becomes important, which we have ignored in our calculations.
Taking account of the periodicity of momentum,
one finds that
coordinate operator $x$ should have discrete eigenvalues.
In order to check that it is possible to deal with this problem,
we have expanded the Hamiltonian
up to second order with respect to the coordinates rather than momenta,
keeping all order in momenta. We 
have also imposed periodic boundary conditions on the wave function
$\Psi(p) = \Psi(p + \pi/2)$.
The result of this quantisation is shown in Fig. \ref{fig:singlej-2}.
The agreement is good in the no-pairing limit  $G=0$, while it is not
as good as the standard quantisation anywhere else.
Since we are not really interested in the (integrable) pure quadrupole
model, but rather in competition between the pairing and quadrupole
forces, we shall ignore the $G=0$ limit in the rest of this work.
We shall thus follow the conventional adiabatic quantisation procedure
in coordinate space as described above.

%%%%%%%%%%%%%%%%%%%%%%%%%%%%%%%%%%%%%%%%%%%%%%%%%%%%%%%%%%%%%%%%%%%%%%%%
\section{The Multi-O(4) Model}
\label{sec:multi_O4}

%%%%%%%%%%%%%%%%%%%%%%%%%%%%%%%%%%%%%%%%%%%%%%%%%%%%%%%%%%%%%%%%%%%%%%%%
\subsection{The Model and the Hamilton's Form of TDH(F)B Equation}

It is a greater challenge to our approach to study the multi-shell
case.  There exists a straightforward extension of the model, by
addition of the individual pairing generators, and summing the
quadrupole operators of each shell with a weight factor (we shall thus
not have a direct coupling between the different shells). The
operators are, for $j=j_1,j_2,\cdots,j_\Lambda$, (for each shell $j_i$
we take the pair degeneracy $\Omega_i = j_i + 1/2$ to be even)
\begin{eqnarray}
P^\dagger &=& \sum_{i,m_i>0} c_{j_im}^\dagger c_{j_i\bar m_i}^\dagger \ ,\quad
\tilde{P}^\dagger =
 \sum_{i,m_i>0} \sigma_{j_im_i} c_{j_im_i}^\dagger c_{j_i\bar m_i}^\dagger \ ,\\
N         &=& \sum_{im_i} c_{j_im_i}^\dagger c_{j_im_i} \ ,\quad
Q         =
 \sum_{im_i} q_i \sigma_{j_im_i} c_{j_im_i}^\dagger c_{j_im_i} \,
\end{eqnarray}
where $q_i$ represents the magnitude of quadrupole moment carried by
single-particle states.
For each shell we can define quasi-spin
$[su(2)\oplus su(2)]$ generators ${\bf A}^i$ and ${\bf B}^i$
in a manner similar to Eqs.\ (\ref{su2_gen_1}) and (\ref{su2_gen_2}).
We choose a slightly more general Hamiltonian than in the previous chapter
by adding a term containing spherical single-particle energies,
\begin{equation}
H = \sum_{jm} \epsilon_j c_{jm}^\dagger c_{jm}
     -G P^\dagger P - \frac{1}{2} \kappa Q^2 \ .
\end{equation}
The exact solution can be also obtained by diagonalisation in a basis set
\begin{equation}
\bigotimes_{i=1}^\Lambda \ket{k_a^{(i)}, k_b^{(i)}}
= \prod_{i=1}^\Lambda (A_+^i)^{k_a^i} (B_+^i)^{k_b^i}
           \ket{0} \ ,
\end{equation}
where $0 \leq k_a^i,k_b^i \leq \Omega_i/2$ and $\sum_i (k_a^i + k_b^i)
= n_0 = N_0/2$.  This is no longer as trivial a calculation as before,
since the dimension of the basis increases rapidly with the number of
shells $\Lambda$, but can still be done, provided that one chooses
$\Lambda$ sufficiently small.  On the other hand, since the
dimension of TDH(F)B configuration space increases only linearly with
$\Lambda$, the amount of effort required for the ALACM calculation is
still rather small.  The time-dependent mean-field state is obtained through
the use of the coherent-state representation as before,
and is given by
\begin{eqnarray}
\ket{\bf z} &=& \exp\left[\sum_{i=1}^\Lambda z^i_a A^i_+ - z_a^{i*} A^i_-
                          + z^i_b B^i_+ - z^{i*}_b B^i_- \right] \ket{0} \ ,\\
   &=& \prod_i^\Lambda
       \left( \cos\frac{\theta_i}{2} \cos\frac{\chi_i}{2} \right)^{\Omega_i/2}
       \exp\left[ \sum_i \tan\frac{\theta_i}{2} \exp(-i\phi_i) A^i_+
             + \tan\frac{\chi_i}{2} \exp(-i\psi_i) B^i_+ \right] \ .
\end{eqnarray}
A possible choice of  canonical variables is
\begin{eqnarray}
\xi^\alpha &=& \left\{ \begin{array}{ll}
               \frac{\Omega_i}{2} \sin^2 \frac{\theta_i}{2}\ , &
               \mbox{for } \alpha = i = 1,\cdots,\Lambda \ ,\\
               \frac{\Omega_i}{2} \sin^2 \frac{\chi_i}{2}\ ,  &
               \mbox{for } \alpha = \Lambda+i = \Lambda+1,\cdots,2\Lambda \ ,
               \end{array} \right. \\
\pi_\alpha &=& \left\{ \begin{array}{ll}
  -\phi_i\ , & \mbox{for } \alpha = i = 1,\cdots,\Lambda \ ,\\
  -\psi_i\ , & \mbox{for } \alpha = \Lambda+i = \Lambda+1,\cdots,2\Lambda \ .
               \end{array}\right.
\end{eqnarray}
It is convenient to allow the indices of $e$, $q$ and $\Omega$ to range
from 1to $2\Lambda$ by copying the original list
of parameters, e.g.,
\begin{equation}
e_\alpha = \left\{ \begin{array}{ll}
        e_i\quad(i=\alpha)  &
               \mbox{for } \alpha =  1,\cdots,\Lambda \ ,\\
        e_i\quad(i=\alpha-\Lambda)   &
               \mbox{for } \alpha = \Lambda+1,\cdots,2\Lambda \ .
               \end{array} \right.
\end{equation}
Using these definitions, the classical Hamiltonian can be given in the compact
form
\begin{eqnarray}
\Ham &=& {h}_{\rm sp} + {\cal H}_P + {\cal H}_Q \ ,\\
\label{clH_P_2}
\Ham_P(\xi,\pi) &=& -\frac{G}{16} \left\{
               \left| \sum_\alpha e^{i\pi_\alpha} S_\alpha \right|^2
              + 32 \sum_\alpha \Omega_\alpha^{-1} (\xi^\alpha)^2 \right\} \ ,\\
\label{clH_Q_2}
\Ham_Q(\xi) &=& -2\kappa \left\{
    \left( \sum_\alpha \sigma_\alpha q_\alpha \xi_\alpha \right)^2
   +  \sum_\alpha \Omega_\alpha^{-1} q_\alpha \xi^\alpha
          (\Omega_\alpha - \xi^\alpha ) \right\} \ ,\\
S_\alpha &=& 2 \sqrt{2\xi^\alpha(\Omega_\alpha - 2\xi^\alpha)} \ ,\\
\sigma_\alpha &=& \left\{ \begin{array}{l}
             +1 \quad \mbox{for } \alpha = 1,\cdots,\Lambda \ ,\\
             -1 \quad \mbox{for } \alpha = \Lambda+1,\cdots,2\Lambda \ .
             \end{array} \right.
\end{eqnarray}
The adiabatic limit of this Hamiltonian is
\begin{eqnarray}
\Ham_{\rm ad} &=& \frac{1}{2} \sum_{\alpha\beta}
                      B^{\alpha\beta} \pi_\alpha \pi_\beta
                      + V(\xi) \ ,\\
B^{\alpha\beta} &=& \frac{G}{8} 
\left[\delta_{\alpha\beta}
       \left( S_\alpha\sum_\gamma S_\gamma\right) -S_\alpha S_\beta \right] \ ,\\
V(\xi) &=& V_P(\xi) + V_Q(\xi) \ ,\\
V_P(\xi) &=& \Ham_P(\xi, \pi=0) \ , \quad V_Q(\xi) = \Ham_Q (\xi) \ .
\end{eqnarray}
Once again, the terms in Eqs. (\ref{clH_P_2}) and (\ref{clH_Q_2})
proportional to $\Omega_a^{-1}$ arise from the exchange contributions, and
will be neglected.

%%%%%%%%%%%%%%%%%%%%%%%%%%%%%%%%%%%%%%%%%%%%%%%%%%%%%%%%%%%%%%%%%%%%%%%%
\subsection{Results for Transitional Nuclei}

Using the LHA we identify the collective degree
of freedom amongst the  $2\Lambda$ coordinates. As before we first must 
remove the NG mode corresponding to a change in particle number explicitly,
due to the zero mass parameter associated with this mode.
The particle number is simply given by the sum of the numbers for the 
individual shells, ${\cal N} = 2 \sum_\alpha \xi^\alpha$. 
It is easy to show that
\begin{equation}
\sum_\beta B^{\alpha\beta} {\cal N}_{,\beta}
                          = 2 \sum_\beta B^{\alpha\beta} = 0 \ .
\end{equation}
Thus we apply the prescription discussed in
Sec.~\ref{sec:prescription} to this model, and use the LHA to
determine the collective path in the remaining
($2\Lambda-1$)-dimensional coordinate space. Since we shall mainly
investigate how the LHA can deal with the transition spherical to
deformed, it is sufficient to study only two shells. We take the size
of these shells to be equal, $\Omega_1 = \Omega_2 = 10$, and put $16$
particles in the available space. We split the degeneracy by taking
$e_1 = 0$ and $e_2=1$, and we use a different value of the ``quadrupole
moment'' for each shell as well, $q_1 = 3$ and $q_2=1$.  The pairing
force is fixed at $G=0.3$, and we only vary the quadrupole force
strength $\kappa$.

\begin{figure}[htb]
\centerline{\includegraphics[width=0.8\textwidth]{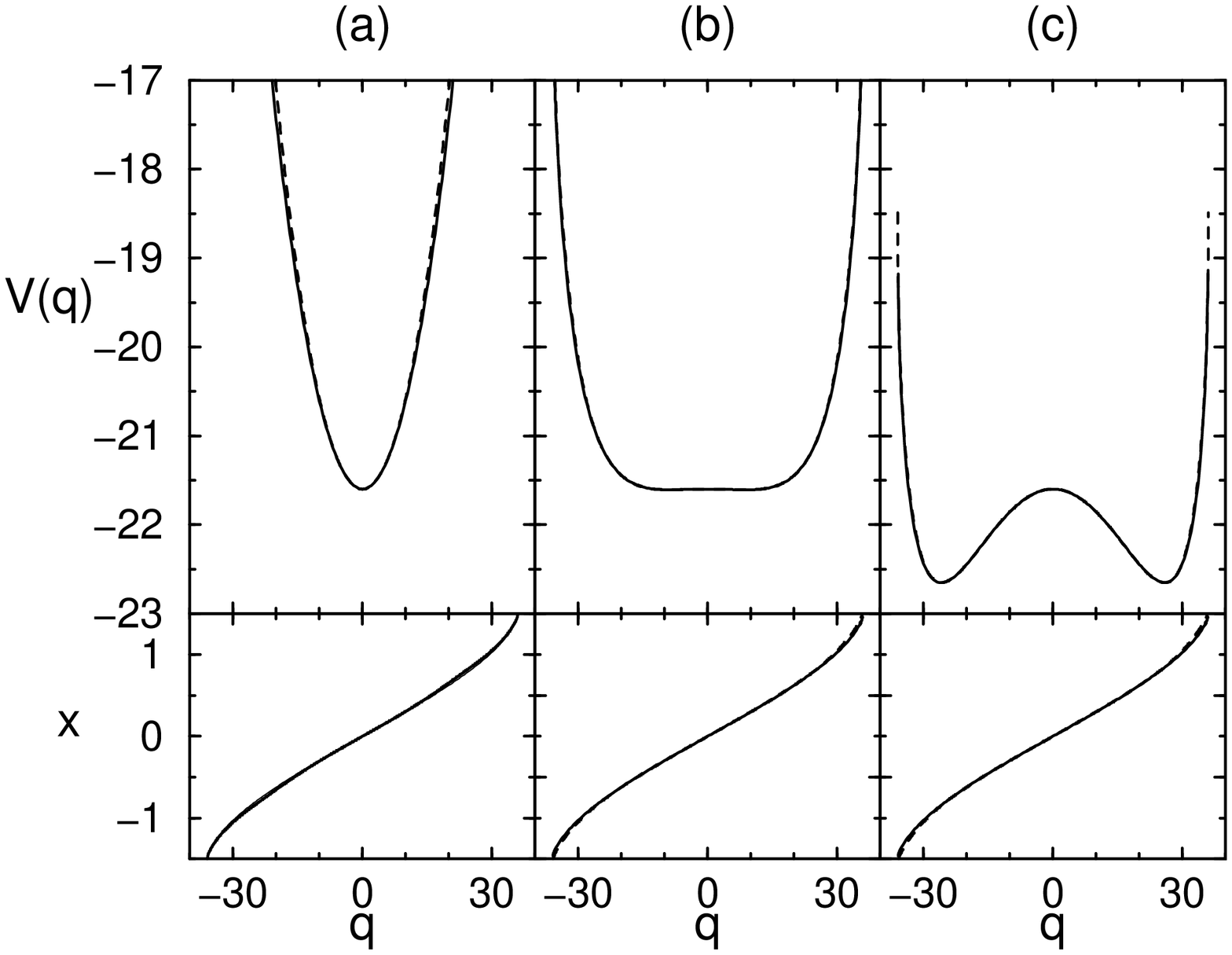}}
\caption{The collective potential energy $V(q)$ and the collective
coordinate $x$ (normalised to unit mass) as a function of the
quadrupole moment $q=\langle \hat Q \rangle$.  We show both the LHA 
(dashed line) and the
CHB (solid line) results in each figure. These results
are, for this model, essentially indistinguishable. The case (a)
corresponds to a weak quadrupole force ($\kappa=0.01$), (b) to a
slightly stronger force ($\kappa=0.03$) and (c) to the strongest,
$\kappa=0.35$. }\label{fig:2shell1}
\end{figure}

We show a representation of the collective potential energy and the
collective coordinate in Fig.~\ref{fig:2shell1}. We represent these
quantities as a function of the expectation value of the quadrupole
operator.  As an alternative to the LHA machinery, we have also
performed a simple constrained Hartree-Bogoliubov (CHB) calculation,
where one imposes  a value for the expectation value of the quadrupole
operator. We determine the mass for this case by replacing the RPA
eigenvector by the coordinate derivative of the quadrupole expectation
value. We then renormalise the coordinate to obtain a constant mass.

\begin{figure}[htb]
\centerline{\includegraphics[width=0.9\textwidth]{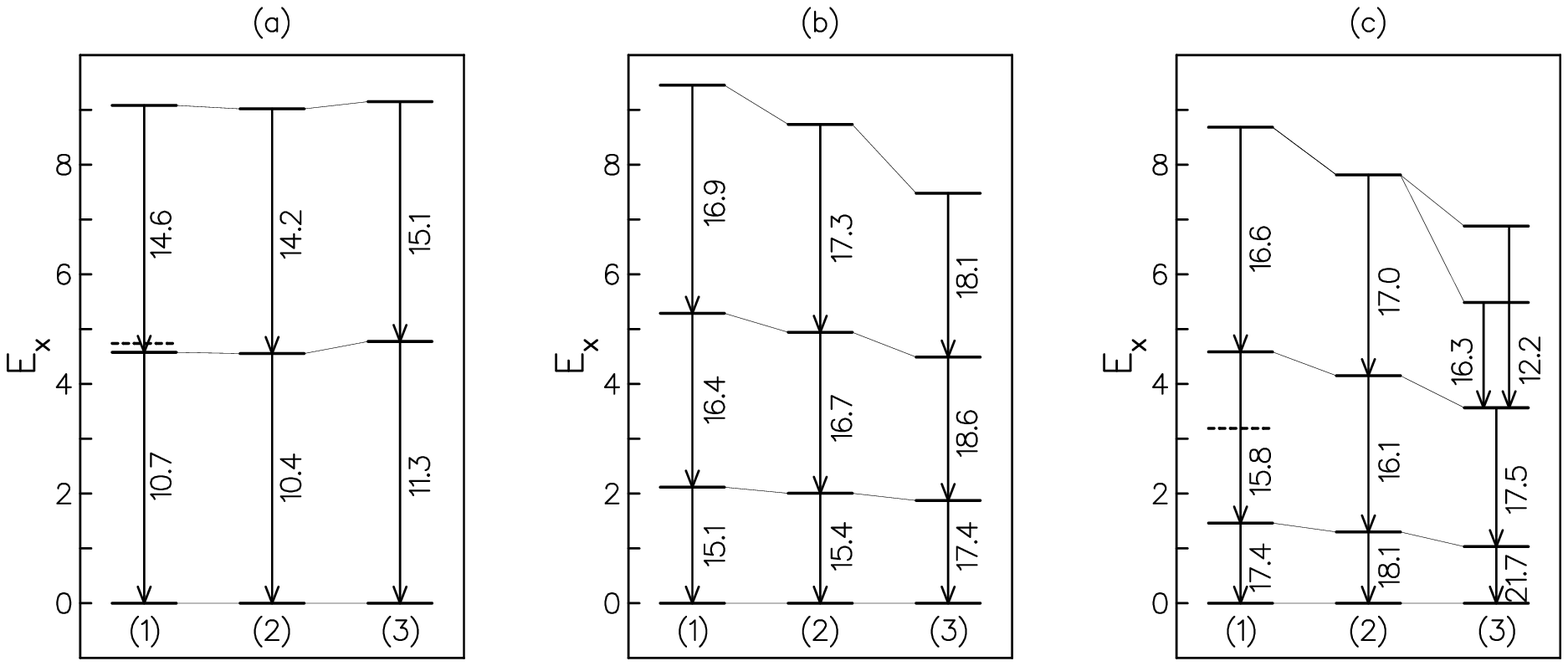}}
\caption{The excitation energies $E_x$ and transition matrix elements
$|\bra{n'} Q \ket{n}|$ (numbers next to arrows) in the two-shell case
discussed in the text.  The three cases (a), (b) and (c) correspond to
those discussed in Fig.~\ref{fig:2shell1}. In each of the three panels
the left one (1) is obtained from requantising the CHB, the middle one
(2) is obtained from requantising the LHA result, and the right one
corresponds to exact diagonalisation. The dashed line shows the lowest
RPA eigenvalue.}
\label{fig:2shell2}
\end{figure}

We have investigated a full shape transition scenario, where we have
changed the strength of the quadrupole interaction so that the collective
Hamiltonian changes from spherical and harmonic for case (a) to flat
for case (b) to deformed for case (c). We see that the difference
between the LHA and CHB calculations is relatively small. This is also
borne out by the spectra and transition strength in
Fig.~\ref{fig:2shell2}. We can see the similarity between the two
approximate calculations. If anything, the LHA gives slightly better
results than the CHB based calculations. We seem to be unable to
reproduce the large density of states found in the exact calculation
for ``deformed'' nuclei, where there are indications from the
transition strengths that some of the states in the approximate
calculations are split into several of the exact states. Note,
however, that at an excitation energy of 6, we are 5 units above the
barrier, so this may just be due to our choice of parameters. The
shape mixing in the low-lying excited states appears to be described
sensibly, however. We would have liked to be able to choose an even
large value of $\kappa$, but if we do that the system collapses to the
largest possible quadrupole moment in the model space, which leads to
all kinds of unphysical complications.

%%%%%%%%%%%%%%%%%%%%%%%%%%%%%%%%%%%%%%%%%%%%%%%%%%%%%%%%%%%%%%%%%%%%%%%%
\section{A multi-shell $O(4)$ model with neutrons and protons}
\label{sec:two_comp_O4}

In heavy nuclei, the  number of neutrons is normally (much) 
larger than that of protons, which often leads to a radically
different shell structure at the Fermi surface for neutrons and protons.
In order to perform a model study of such phenomena, where we can
still perform an exact calculation, we adapt 
the multi-shell $O(4)$ model introduced in the previous section
to one describing systems with both neutrons and protons
\cite{FMM91}.
 We shall
then use this model to analyse the collective dynamics of 
shape-coexistence nuclei, as observed for instance in semi-magic nuclei.
At the same time we shall concentrate on the diabatic/adiabatic dichotomy
already mentioned in the introduction.

The model is a simple extension of the multi-shell $O(4)$ model
in the previous section, with the main difference that we do not have
pairing between proton and neutron orbitals,
\begin{eqnarray}
H &=& H_n + H_p + H_{np}\ ,\\
H_n &=& \sum_{i\in n,m_i} \epsilon_i c_{j_im_i}^\dagger c_{j_im_i}
       -G_n P_n^\dagger P_n -\frac{1}{2}\kappa Q_n^2 \ ,\\
H_p &=& \sum_{i\in p,m_i} \epsilon_i c_{j_im_i}^\dagger c_{j_im_i}
       -G_p P_p^\dagger P_p -\frac{1}{2}\kappa Q_p^2 \ ,\\
H_{np} &=& - \kappa Q_n Q_p \ ,
\end{eqnarray}
where $P_{n(p)}$ and $Q_{n(p)}$ are the pairing and quadrupole operators
for neutrons (protons) (see the definitions in Sec.~\ref{sec:multi_O4}).
In this model there are two trivial NG modes associated with
the change of neutron and of proton number, which can both be removed 
explicitly in the manner discussed before. 

We study this model for a single-shell for neutrons, with pair
degeneracy $\Omega_n=50$, containing 40 particles. We take the
single-particle quadrupole matrix element $q_n=1$, and use a pairing
strength $G_n=0.3$, and assume zero single-particle energy. For
protons we take two shells, both with $\Omega_{p1}=\Omega_{p2}=2$,
$q_{p1}, q_{p2}=2$, having single-particle energies
$e_{p1}=-e_{p2}=5$. We study two different set of interaction
parameters, both with $\kappa=0.1$. The first is $G_n=G_p=0.3$, the
second has the same neutron pairing strength, but $G_p=10$.

\begin{figure}[htb]
\centerline{\includegraphics[width=0.8\textwidth]{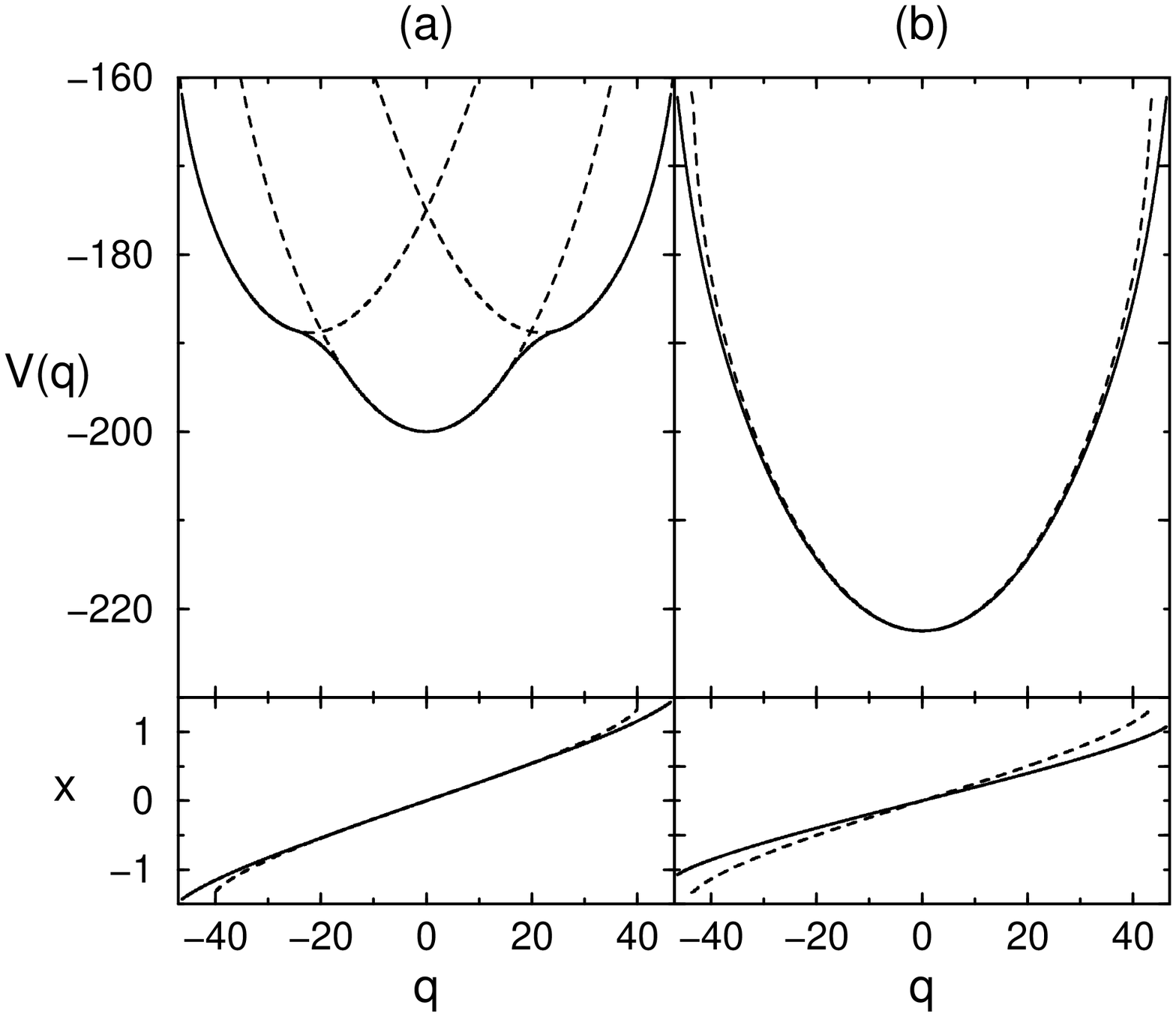}}
\caption{The collective potential energy $V(q)$ and the collective coordinate
$x$ as a function of the quadrupole moment $q= q_p+q_n $.  We
show both the LHA (dashed lines) and the CHB
results (solid line) in
each figure.  The case (a) corresponds to a weak proton
pairing force ($G_p=0.3$), (b) to a strong pairing force ($G_p=10$).
The rest of the parameters are given in the main text.
}\label{fig:pnshell1}
\end{figure}

The collective potential energy for the weaker pairing strength shows
a very interesting behaviour, with two shoulders appearing in the CHB
collective potential energy. This is what is normally called the
adiabatic potential energy, and the shoulders arise from an avoided
crossing. As in our previous work \cite{NW98} the adiabatic LHA method
chooses a diabatic (crossing) set of potential energy curves. These
shape-coexistence minima are related to 2p-2h excitation in the proton model
space, promoting two particles from the lowest Nilsson orbitals to the
down-sloping ones. This is of course very similar to the phenomena
observed in shape coexistence in semi-magic nuclei. 

We get another surprise for the strong pairing case. Here the
collective potentials look very similar, and rather structureless, but
the collective coordinates are different. This can be traced back to
the fact that the collective coordinate in the LHA is not $q_p+q_n$,
but a different combination. At the minimum, the lowest RPA mode
correspond approximately to $q_n+\frac{1}{10}q_p$.  This is similar to
the situation analysed in great detail in Ref.~\cite{DWK94}, and once
again shows the importance of self-consistency in the selection of the
collective coordinate. The real collective coordinate is {\em not} the
mass-quadrupole operator!

\begin{figure}[htb]
\centerline{\includegraphics[width=0.9\textwidth]{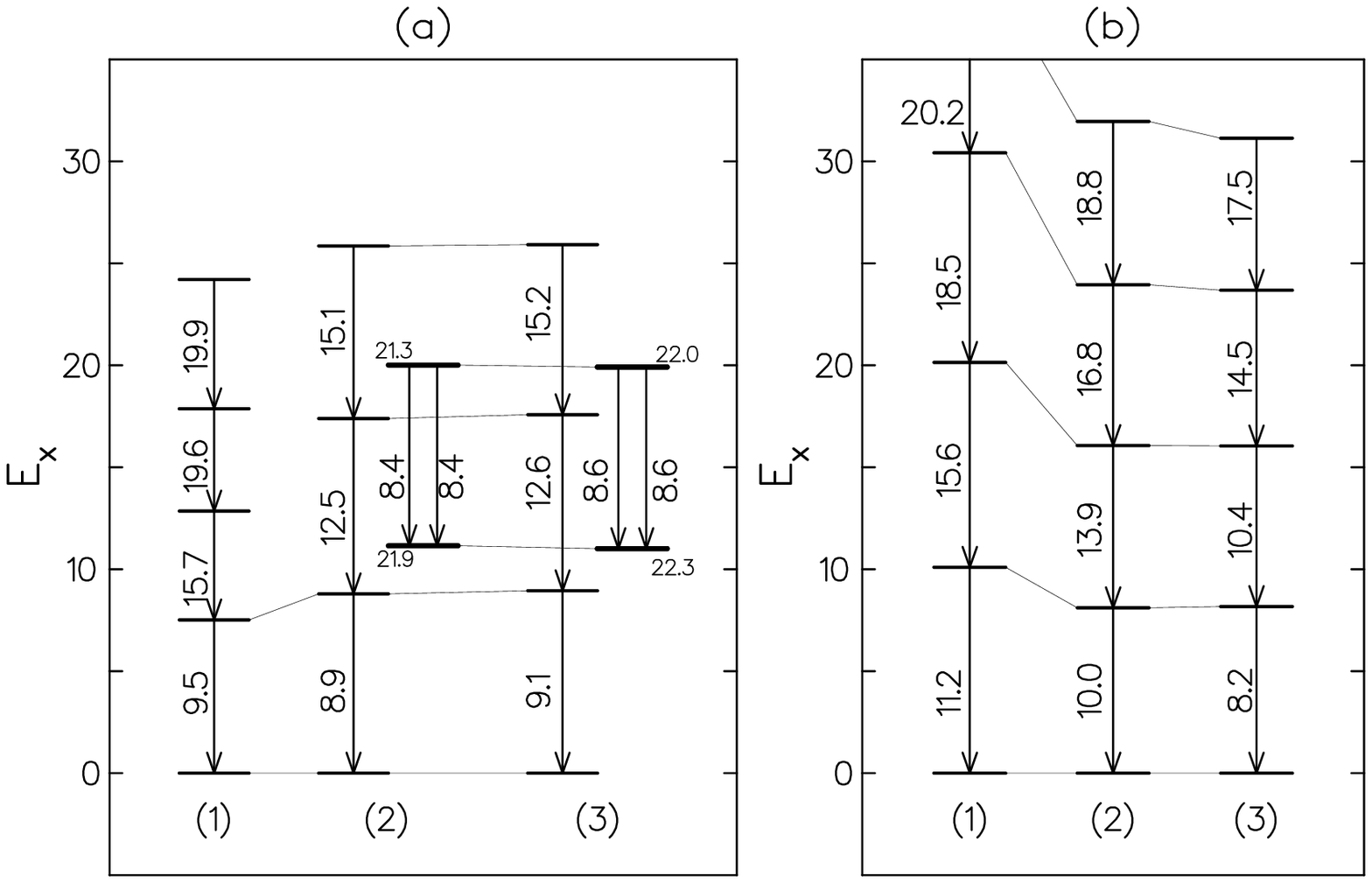}}
\caption{The excitation energies $E_x$ and transition matrix elements
$|\bra{n'} Q \ket{n}|$ (numbers next to arrows) in the proton-neutron
case discussed in the text.  The two cases (a) and (b) correspond to
those discussed in Fig.~\ref{fig:pnshell1}. In each of the panels the
left one (1) is obtained from requantising the CHB, the middle one (2)
is obtained from requantising the LHA result, and the right one (3)
corresponds to the exact diagonalisation. Levels denoted by thick
lines are doubly degenerate.  The numbers next to the arrows denote
teh size of the transition matrix elements.}
\label{fig:pnshell2}
\end{figure}

In Fig.~\ref{fig:pnshell2} we show the consequences of these results 
for the requantisation. For the weak pairing case the diabatic picture
obtained through the LHA gives almost perfect results, whereas the CHB
potential energy curve fails to provide the correct answer. In the case of
the strong pairing the incorrect choice of the collective coordinate
leads to too large a level spacing in the CHB calculations, whereas the LHA
and exact calculations agree again.

For the weak pairing case the exact calculation almost exhibits the doublet
structure found from the diabatic potential energy curves; the splitting
is less than one part in $10^5$ in the exact calculation. Nevertheless,
the exact calculation consists of symmetric and antisymmetric states, which 
leads to the transition matrix element 22.3 between these two states. Since
in the LHA calculation the states do not mix, we have printed the
{\em diagonal} matrix element instead. For very weak mixing this is
the correct comparison, as is borne out by the results.

The decoupling measure $D$, Eq.~(\ref{decoupling}), is small for all
these cases. The worst case is the strong pairing case, where it rises
from $0$ for $q=0$ to $3\times10^{-3}$ for $q=40$. For this reason we
also believe that the ``scalar Berry potential'' \cite{NW98} will be
small, and we have not included this, or any other quantum corrections,
in our calculations.

%%%%%%%%%%%%%%%%%%%%%%%%%%%%%%%%%%%%%%%%%%%%%%%%%%%%%%%%%%%%%%%%%%%%%%%%
\section{Conclusion}
\label{sec:conclusion}

In the area of shape coexistence, as studied by the model calculations
discussed in this paper, it seems that on the whole the LHA is an
excellent method to obtain sensible results, with reasonable
effort. Even though the size of the space is much smaller than that
used in a straightforward diagonalisation, the use of the local RPA,
which needs to be performed many times at every point along the
collective path, may seem prohibitive for realistic calculations.
Fortunately, many problems can be studied with simplified separable
forces, as in the model discussed here. We are at the moment
considering the old pairing-plus-quadrupole model, that has been
applied so successfully in the physics of heavy nuclei. Such a model
can be dealt with much more straightforwardly than more microscopic
Skyrme or Gogny-force based approaches. This will allow us to shed
light on a good treatment of shape-coexistence, as well on the
interesting question on the choice of the collective (cranking)
operator, which was already found to be non-trivial in certain limits
of the $O(4)$ model.

One might argue that even that is not enough, and we should really
adopt the full horror of standard nuclear many-body approaches. We
believe that we can address this problem, and are actively considering
the approaches available to us. As must be obvious from the discussion
given above, an efficient calculation hinges on efficient solution of
the RPA. We are investigating two approaches to this problem, the use
of iterative diagonalisation of the RPA using Lanczos procedures, or
the approximation of the RPA by  using separable forces
\cite{sep:RPA}, which can be diagonalised much more efficiently.

In the present work we have not included any quantum corrections, nor
the ``scalar Berry potential'' arising from choosing a slow degree of
freedom.  In our previous work \cite{NW98} we have shown that these
corrections tend to improve the spectrum, so that the results
presented can even be improved slightly.  We are strongly encouraged
by the ability of our method to choose the right coordinate and the
correct potential energy curves for problems that involve pairing and
``quadrupole'' degrees of freedom, and we expect results for realistic
nuclear models in the near future.

\acknowledgements

This work was supported by a research grant (GR/L22331) from
the Engineering and Physical Sciences Research Council (EPSRC)
of Great Britain.

%%%%%%%%%%%%%%%%%%%%%%%%%%%%%%%%%%%%%%%%%%%%%%%%%%%%%%%%%%%%%%%%%%%%%%%%

%%%%%%%%%%%%%%%%%%%%%%%%%%%%%%%%%%%%%%%%%%%%%%%%%%%%%%%%%%%%%%%%%%%%%%%%


\begin{thebibliography}{99}
\bibitem{KWD91}
A.~Klein, N.R.~Walet and G.~Do~Dang, Ann. Phys. {\bf 208}, 90 (1991).
\bibitem{NW98} T.~Nakatsukasa and N.R.~Walet, Phys. Rev. {\bf C57},
1192 (1998).
\bibitem{PSS80}
R.~Piepenbring, B. Silvestre-Brac and Z.~Szymanski, Nucl. Phys. {\bf A348},
77 (1980).
\bibitem{Mat82}
K.~Matsuyanagi, Prog. Theor. Phys. {\bf 67}, 1441 (1982);
Proceedings of the Nuclear Physics Workshop, Trieste, 5-30 Oct. 1981.
ed. C.H. Dasso, R.A. Broglia and A. Winther (North-Holland, 1982),
p.29. 
\bibitem{Mat86}
M.~Matsuo, Prog. Theor. Phys. {\bf 76}, 372 (1986).
\bibitem{SM88}
T.~Suzuki and Y.~Mizobuchi, Prog. Theor. Phys. {\bf 79}, 480 (1988);
Y.~Mizobuchi, Prog. Theor. Phys. {\bf 65}, 1450 (1981).
\bibitem{FMM91}
T.~Fukui, M.~Matsuo and K.~Matsuyanagi,
Prog. Theor. Phys. {\bf 85}, 281 (1991).
\bibitem{AB88}
P.O.~Arve and G.F.~Bertsch, Phys. Lett. {\bf B215}, 1 (1988).
\bibitem{PPQQ} D.R. B\`es and R.A. Sorensen, Adv. Nucl. Phys. {\bf 2}
(1969) 129.
\bibitem{KT72}
T. Kishimoto and T. Tamura, Nucl. Phys. {\bf A192}, 246, (1972);
Nucl. Phys. {\bf A270}, 317, (1976).
\bibitem{Woo92}
J.L.~Wood, K.~Heyde, W.~Nazarewicz, M.~Huyse and P.~van~Duppen,
Phys. Rep. {\bf 215}, 101 (1992).
\bibitem{BN89}
R.~Bengtsson and W.~Nazarewicz, Z. Phys. {\bf A334}, 269 (1989).
\bibitem{HW53}
D.L.~Hill and J.A.~Wheeler, Phys. Rev. {\bf 89}, 1102 (1953).
\bibitem{GR} P.G.~Reinhard and K.~Goeke, Rep. Prog. Phys. {\bf 50}, 1 (1987).
\bibitem{Per86}
A.~Perelomov, {\it Generalized Coherent States and Their Applications},
(Springer-Verlag, Berlin, 1986).
\bibitem{DWK94} G. Do Dang, N.R. Walet, and A. Klein, Phys. Lett. 
{\bf B322}, 11 (1994).
\bibitem{sep:RPA} V.O. Nesterenko, W. Kleinig, V.V. Gudkov, and J. Kvasil,
Phys. Rev. C {\bf 53}, 1632 (1996).

\end{thebibliography}
\end{document}